\documentclass{ws-procs9x6-cpt25}
\begin{document}
	
	\newcommand{\refeq}[1]{(\ref{#1})}
	\def\etal {{\it et al.}}
	
	\title{Tests of Lorentz Symmetry using X-ray Polarimetry}
	
	\author{F.\ Kislat}
	
	\address{Department of Physics \& Astronomy and Space Science Center,\\
		University of New Hampshire,\\
		Durham, NH 03824, USA}
	
	\begin{abstract}
		Lorentz symmetry is the fundamental symmetry of Einstein's theory of Special Relativity and has been tested to great precision.
		Nevertheless, the possibility remains that it is violated at the Planck scale, as predicted by some theories of quantum gravity.
		While the Planck scale is not directly accessible to experiments, minute residual deviations from Lorentz symmetry at attainable energies may be observable.
		The polarization of light from astrophysical sources is a particularly powerful probe because tiny differences accumulate as light travels over astrophysical distances, and polarization is sensitive to light travel time differences between polarization modes on the order of the oscillation period of the electromagnetic wave.
		Here, we report on new constraints on Lorentz invariance violation derived from X-ray polarization measurements of active galactic nuclei.
		The new constraints, presented in the framework of the Standard-Model Extension, improve on our previous work, which used optical polarization measurements, by four orders of magnitude.
	\end{abstract}
	
	\bodymatter
	
	\section{Introduction}
	Although Lorentz invariance has been tested to great precision,\cite{datatables} some theories of quantum gravity indicate that the symmetry may be broken at the Planck scale.\cite{liv}
	Residual effects may exist at attainable energies, but numerous experiments have established that, if present, these effects must be tiny.
	As a consequence of Lorentz symmetry, the speed of light is the same in all reference frames, invariant under boosts and rotations.
	This implies that the speed of light must be independent of wavelength (or photon energy), propagation direction, and polarization.
	
	When photons propagate over astronomical distances, even minuscule deviations from the speed of light, $c$, will accumulate and may result in observable effects.
	As a result, some of the strongest constraints on Lorentz invariance violations are due to astrophysical and cosmological observations.\cite{datatables}
	The greatest challenge of such tests is that the source of the observed light cannot be controlled and details of the emission are generally unknown.
	Nevertheless, robust constraints on Lorentz-violating theories can be derived from observations by making a few conservative assumptions, illustrated most easily with the help of two examples.
	
	Explosive events such as gamma-ray bursts (GRBs) and supernova explosions result in the emission of large amounts of radiation on timescales as short as seconds.
	While the absolute propagation time cannot be measured, it is possible to compare the arrival time of photons at different energies, enabling a test of the photon dispersion relation.\cite{dispersion}
	For a typical GRB, arrival time differences can be constrained to $\mathcal{O}(1\,\mathrm{s})$.
	In general, such measurements are limited by photon statistics, not the timing precision of the instrument.
	If the GRB is at redshift $z=1$ (light travel time ${\sim}7.7\times 10^{12}\,\mathrm{a}$), this amounts to a sensitivity to differences in propagation velocity of $\mathcal{O}(10^{-20})$.
	Of course, this estimate implies the assumption that light of all wavelengths is emitted at the same time.
	That need not be true, but conservative tests can be constructed in such a way that reliable upper limits on velocity differences can be derived except in the unlikely case that astrophysical effects and Lorentz-invariance-violating effects cancel.
	
	Even tighter constraints can be obtained from polarization measurements.\cite{kaaret_2004,kostelecky_mewes_2013,kostelecky_mewes_2009,optical_polarimetry}
	Light travel time differences between different polarization modes on the order of the oscillation period of the wave, $\nu^{-1}$, will result in a measurable change of the polarization angle compared to the source.
	The absence of such a change then results in a constraint on birefringence that is a factor $\mathcal{O}(\nu^{-1})$ stronger than time-of-flight constraints from the same source.
	Although the polarization at the source is unknown, any energy dependence of such an effect will lead to an energy-dependent observed polarization even if the polarization at the source does not depend on energy.
	The assumption of energy-independent polarization at the source can often be motivated if the emission mechanism is the same across the observational energy range.
	Even if this is not the case, a non-detection of energy dependence can be used to derive constraints on birefringence under the assumption that it is very unlikely that astrophysical and Lorentz-violating effects will cancel.
	
	The Standard-Model Extension (SME) is an effective field theory framework to describe Lorentz- and CPT-invariance-violating effects in theories beyond the Standard Model.\cite{sme,kostelecky_mewes_2009}
	The effect of each of the higher-order operators is described by a set of coefficients, which can be ordered by the mass-dimension $d$ of the corresponding operator.
	Working in this framework has the advantage that it enables a direct comparison of results obtained from different experimental approaches and to translate those results into constraints on the underlying theories.
	
	Previously, we have used optical polarimetry of active galactic nuclei (AGN) and GRB afterglows to derive some of the strongest constraints on the $d=4,5,6$ coefficients of the SME photon sector.\cite{optical_polarimetry,friedman_etal_2020}
	Here, we extend this approach to include X-ray polarimetry results from the Imaging X-ray Polarimetry Explorer\cite{weisskopf_etal_2022} mission (IXPE) and derive new constraints on the $d=5$ isotropic model.
	
	The remainder of the paper is structured as follows: We describe our methodology in Section~\ref{sec:methods}, present the results in Section~\ref{sec:results}, and finally give a summary and outlook towards future work in Section~\ref{sec:summary}.
	
	\section{Methods}\label{sec:methods}
	To explain our approach, we begin with a brief review of the relevant theory following Ref.~\refcite{kostelecky_mewes_2009}.
	In the SME, Lorentz-violating operators result in a modified photon dispersion relation,
	\begin{equation}
		E \simeq \left(1 - \varsigma^0 \pm \sqrt{(\varsigma^1)^2 + (\varsigma^2)^2 + (\varsigma^3)^2}\right)p,
	\end{equation}
	where the terms $\varsigma^i$ depend on photon energy and propagation direction, determined by the SME coefficients, and the two signs correspond to left- and right-handed polarization.
	The terms $\varsigma^0$, $\varsigma^1$, and $\varsigma^2$ only depend on coefficients of even mass dimension, and $\varsigma^3$ only depends on odd-mass-dimension coefficients,
	\begin{equation}
		\varsigma^3 = \sum_{\substack{d\text{ odd}\\jm}}E^{d-4}\,{}_0Y_{jm}(\hat{\boldsymbol{n}})k_{(V)jm}^{(d)}.
	\end{equation}
	Here, ${}_0Y_{jm}$ are spherical harmonics, $\hat{\boldsymbol{n}}$ is a unit vector pointing in the opposite direction of the photon propagation (i.e.,\ toward the source), and $k_{(V)jm}^{(d)}$ are the SME coefficients.
	Not all values of $j,m$ are allowed and at $d = 5$ there are 16 coefficients in total.
	However, in this paper we will only consider the $d=5$ isotropic term,
	\begin{equation}
		\varsigma^3_{(5)\text{iso}} = E\,{}_0Y_{00}k_{(V)00}^{(5)}.
	\end{equation}
	An extension of the analysis presented here to anisotropic models and other mass dimensions will be the subject of a forthcoming paper.
	
	As a photon propagates, its Stokes vector, $\boldsymbol{s} = (Q,U,V)^T$, will rotate around an axis $\boldsymbol{\varsigma} = (\varsigma^1, \varsigma^2, \varsigma^3)^T$ determined by the terms in the dispersion relation:
	\begin{equation}
		\frac{d\boldsymbol{s}}{dt} = 2E\boldsymbol{\varsigma}\times\boldsymbol{s}.
	\end{equation}
	Because $\varsigma^1$ and $\varsigma^2$ vanish for any odd $d$, the birefringence axis $\boldsymbol{\varsigma}$ points in the $V$ direction, which means that circular polarization $V$ will be unaffected while the linear polarization direction $\psi$ rotates continuously.
	
	Assuming two photons of energies $E_1$ and $E_2$ emitted by a source at redshift $z$ have the same polarization angle at the source, the observed polarization angles at Earth will differ by
	\begin{equation}
		\Delta\psi = (E_1^2-E_2^2)\,{}_0Y_{00}k_{(V)00}^{(5)}\int_0^z\frac{1 + z'}{H_{z'}}dz'
	\end{equation}
	in the $d=5$ isotropic model, where $H_{z'}$ is the Hubble expansion rate at redshift $z'$.
	It is clear from this equation that constraints on $k_{(V)00}^{(5)}$ can be derived by measuring $\Delta\psi$ over some energy range.
	However, in this paper, we take an alternative approach that will allow us to make use of observations that do not allow energy-resolved polarimetry.
	
	Following the approach developed in Refs.~\refcite{optical_polarimetry} and~\refcite{friedman_etal_2020}, we continue to assume that the polarization angle at the source is independent of energy.
	The Stokes parameters of a photon of energy $E_1 \leq E \leq E_2$ within the band pass of IXPE of $E_1 = 2\,\mathrm{keV}$ and $E_2 = 8\,\mathrm{keV}$ at the observer are
	\begin{align}
		q(E) &= \cos\left(2\zeta_{k00}^{(5)}(E^2 - E_1^2)\right), \\
		u(E) &= \sin\left(2\zeta_{k00}^{(5)}(E^2 - E_1^2)\right),
	\end{align}
	where we choose our coordinate system such that $Q(E_1) = U(E_1) = 0$ and define
	\begin{equation}
		\zeta_{k00}^{(5)} = {}_0Y_{00}k_{(V)00}^{(5)}\int_0^z\frac{1 + z'}{H_{z'}}dz'.
	\end{equation}
	If $\Pi(E)$ is the polarization at the source, the spectrally integrated Stokes parameters of an IXPE observation are
	\begin{align}
		Q &= \int_{E_1}^{E_2}\Pi(E)\cos\left(2\zeta_{k00}^{(5)}(E^2 - E_1^2)\right)\mu(E)F(E)\,dE, \\
		U &= \int_{E_1}^{E_2}\Pi(E)\sin\left(2\zeta_{k00}^{(5)}(E^2 - E_1^2)\right)\mu(E)F(E)\,dE,
	\end{align}
	where $F(E)$ is the observed flux at energy $E$, and $\mu(E)$ is the modulation factor of IXPE, which describes the polarization sensitivity of the instrument.
	The modulation factor is defined as the polarization measured for a fully polarized incident beam.
	It is easy to see that in general $\zeta_{k00}^{(5)} \neq 0$, corresponding to $k_{(V)00}^{(5)} \neq 0$, will result in a reduced observed polarization fraction $P = \sqrt{Q^2 + U^2}$ compared to the case of no Lorentz invariance violation.
	
	\begin{figure}[t]
		\centering
		\includegraphics[width=2.8in]{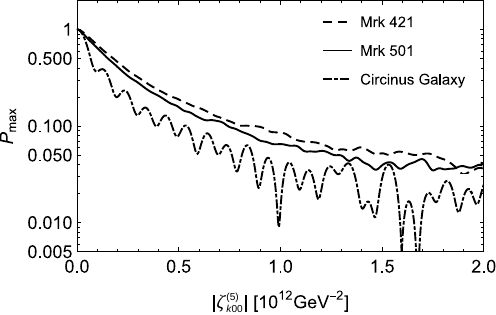}
		\caption{%
			Theoretically possible maximum observable polarization $P_\text{max}$ according to Eq.~\refeq{eq:Pmax} as a function of the Lorentz-violating parameter $\zeta_{k00}^{(5)}$ for three AGN polarization measurements with IXPE.
			The curves are determined primarily by the measured source spectra $F(E)$.%
		}
		\label{fig:Pmax}
	\end{figure}
	
	Because $\Pi(E)$ at the source is unknown, we instead assume $\Pi(E) \equiv 1$ to calculate $Q_\text{max}$ and $U_\text{max}$, which lead to the largest observable polarization
	\begin{equation}\label{eq:Pmax}
		P_\text{max} = \sqrt{Q_\text{max}^2 + U_\text{max}^2}
	\end{equation}
	for a given value of $k_{(V)00}^{(5)}$.
	Any lower polarization at the source will result in a lower observed polarization.
	Figure~\ref{fig:Pmax} shows the maximum observable polarization as a function of $\zeta_{k00}^{(5)}$ for a sample of three IXPE polarization observations of AGNs.
	We then calculate the likelihood $\mathcal{P}(P<P_\text{max}|\sigma_P)$ that the true polarization value at the observer $P$ is less than $P_\text{max}$ given the measured polarization and its uncertainty $\sigma_P$.
	
\begin{figure}
		\centering
		\includegraphics[width=2.8in]{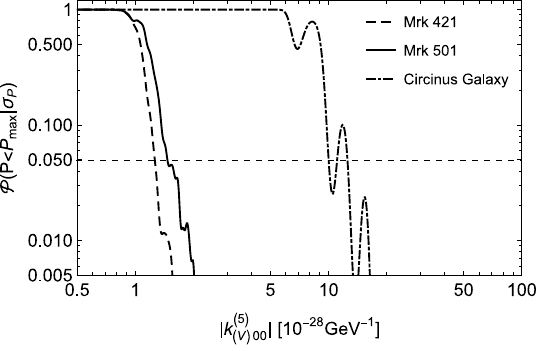}
		\caption{%
			By comparing IXPE polarization measurements (Mrk 421: $(15\pm 2)\%$,\cite{digesu_etal_2022} Mrk 501: $(10\pm 2)\%$,\cite{liodakis_etal_2022} Cir Galaxy: $(20 \pm 3.8)\%$\cite{ursini_etal_2022}) to the maximum observable polarization we calculate the likelihood that the true polarization at Earth is less than the maximum possible polarization, $P_\text{max}$, shown in Fig.~\ref{fig:Pmax}.
			The limit on the SME coefficient $k_{(V)00}^{(5)}$ is then the largest value where $\mathcal{P}(P < P_\text{max}|\sigma_P) \geq 0.05$.%
		}
		\label{fig:Plikelihood}
	\end{figure}

	The resulting likelihood $\mathcal{P}(P<P_\text{max}|\sigma_P)$ for the three AGN observations from Fig.~\ref{fig:Pmax} is shown in Fig.~\ref{fig:Plikelihood}.
	An upper limit on $k_{(V)00}^{(5)}$ at the 95\,\% confidence level is then derived by finding the largest value where $\mathcal{P} \geq 0.05$.
	We find values $k_{(V)00}^{(5)} \lesssim \mathcal{O}(10^{-27}\,\mathrm{GeV}^{-1})$, which is about 3 orders of magnitude smaller than the limits we previously found for an anisotropic model by combining optical polarimetry of 1278 unique astrophysical objects.
	
	Given the $E^2$ dependence of $\Delta\psi$ and the fact that the IXPE energy range is a factor $10^3$ higher than optical frequencies, one might expect a $10^6$ improvement of the result.
	However, our previous results have the advantage that many more optical polarization measurements are available than X-ray polarization measurements, many of which are at higher redshift, and statistically more significant.
	
	\section{Results}\label{sec:results}
	
	\begin{table}[t]
		\tbl{IXPE polarization detections from 11 active galactic nuclei and resulting constraints on the isotropic SME coefficient~$k_{(V)00}^{(5)}$.
			All redshifts were obtained from the SIMBAD astronomical database at \texttt{https://simbad.u-strasbg.fr}.}
		{\begin{tabular}{@{}lcccc@{}}\toprule
				\multicolumn{1}{c}{Source} & Redshift & Polarization & Ref. & $\left|k_{(V)00}^{(5)}\right|_\text{max}$ \\[2ex]
				& $z$ & (\%) & & ($10^{-29}\,\mathrm{GeV}^{-1}$) \\\colrule
				Mrk 421 & 0.030\hphantom{00} & \hphantom{0.}$15 \pm 2$\hphantom{.0} & [\refcite{digesu_etal_2022}] & \hphantom{0}12.5\hphantom{0} \\
				Mrk 501 & 0.033\hphantom{00} & \hphantom{0.}$10 \pm 2$\hphantom{.0} & [\refcite{liodakis_etal_2022}] & \hphantom{0}13.7\hphantom{0} \\
				Cir Galaxy & 0.00145 & \hphantom{0.}$20 \pm 3.8$ & [\refcite{ursini_etal_2022}] & 120\hphantom{.00} \\
				NGC 1068 & 0.00348 & $12.4 \pm 3.6$ & [\refcite{marin_etal_2024}] & 127\hphantom{.00} \\
				NGC 4151 & 0.00315 & \hphantom{0}$4.9 \pm 1.1$ & [\refcite{ngc4151}] & \hphantom{0}73.9\hphantom{0} \\
				H1426+428 & 0.129\hphantom{00} & $20.3 \pm 4.1$ & [\refcite{banerjee_etal_2025}] & \hphantom{00}1.87 \\
				PKS 2155--304 & 0.116\hphantom{00} & $30.7 \pm 2.0$ & [\refcite{kouch_etal_2024}] & \hphantom{00}1.44 \\
				IC 4329A & 0.0160\hphantom{0} & \hphantom{0}$3.3 \pm 1.1$ & [\refcite{ingram_etal_2023}] & \hphantom{0}83.9\hphantom{0} \\
				1ES 1959+650 & 0.047\hphantom{00} & \hphantom{0}$8.0 \pm 2.3$ & [\refcite{errando_etal_2024}] & \hphantom{0}11.7\hphantom{0} \\
				1ES 0229+200 & 0.1397\hphantom{0} & $17.9 \pm 2.8$ & [\refcite{ehlert_etal_2023}] & \hphantom{00}1.83 \\
				PG 1553+113 & 0.360\hphantom{00} & \hphantom{0.}$10 \pm 2$\hphantom{.0} & [\refcite{middei_etal_2023}] & \hphantom{00}1.56 \\\botrule
		\end{tabular}}
		\label{tab:ixpe-results}
	\end{table}
	
	\begin{figure}
		\centering
		\includegraphics[width=3.75in]{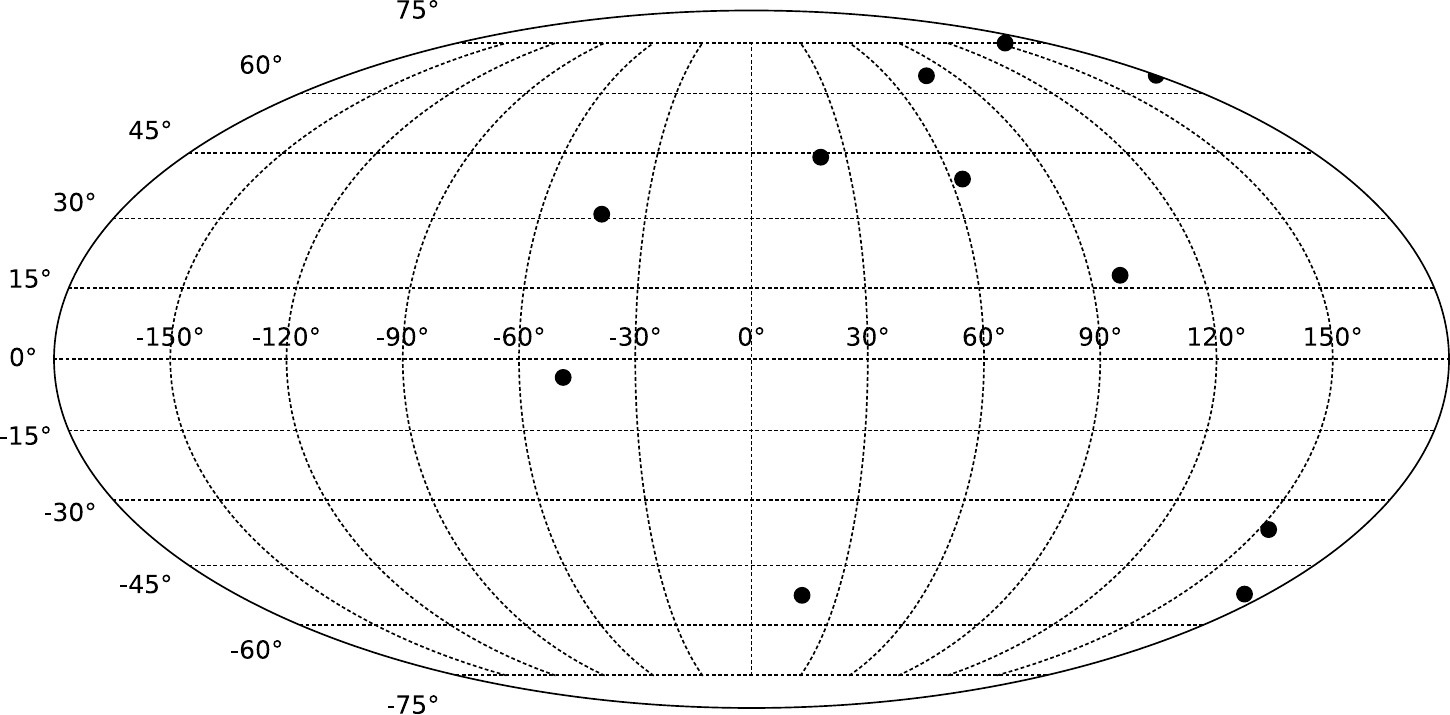}
		\caption{%
			Skymap in galactic coordinates showing the locations of the 11 AGN from which IXPE has detected non-zero polarization to date.%
		}
		\label{fig:skymap}
	\end{figure}
	
	Table~\ref{tab:ixpe-results} lists all IXPE polarization detections from 11 different AGNs and Fig.~\ref{fig:skymap} shows their distribution in the sky.
	It is noteworthy that all are BL Lac-type blazars or Seyfert 1/2 galaxies.
	For low or intermediate-energy peaked blazars, only upper limits on the polarization fraction have been found.
	We applied the methods described in the previous section to each of the sources and the resulting upper limits on $k_{(V)00}^{(5)}$ are also given in the table.
	The strength of the individual constraints is determined primarily by the source redshift $z$ and, to a lesser degree, by the significance of the polarization detection.
	
\begin{figure}
		\centering
		\includegraphics[width=3in]{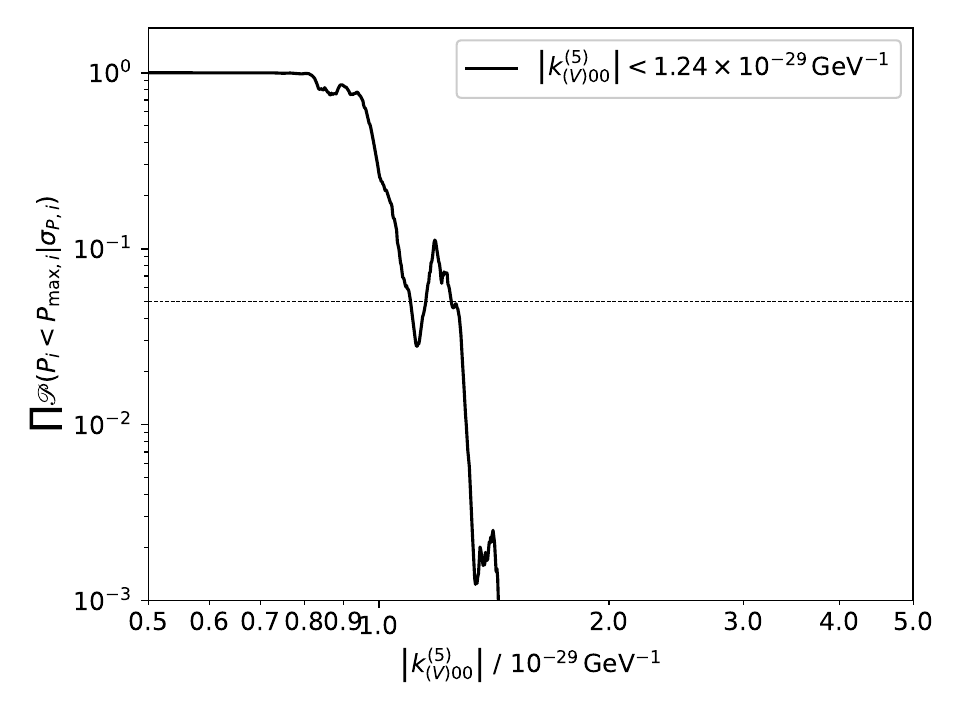}
		\caption{%
			Product likelihood $\prod_i\mathcal{P}(P_i<P_{\text{max},i}|\sigma_{P,i})$ that all IXPE measurements listed in Table~\ref{tab:ixpe-results} are compatible with a given value of $k_{(V)00}^{(5)}$.
			At the 95\,\% confidence level, this leads to the upper limit $\left|k_{(V)00}^{(5)}\right| < 1.24 \times 10^{-29}\,\mathrm{GeV}^{-1}$.%
		}
		\label{fig:product-likelihood}
	\end{figure}

	The individual results can be combined by calculating the product likelihood
	$$
	\prod_i\mathcal{P}(P_i<P_{\text{max},i}|\sigma_{P,i})
	$$
	which is shown in Fig.~\ref{fig:product-likelihood}.
	We find the upper limit
	$$
	\left|k_{(V)00}^{(5)}\right| < 1.24 \times 10^{-29}\,\mathrm{GeV}^{-1},
	$$
	which is dominated by the strongest individual constraint, as expected.
	Although no significant improvement from using multiple sources is found in this isotropic analysis, this approach will make a significant difference in an anisotropic analysis, where all $k_{(V)jm}^{(d)}$ for any given $d$ are constrained simultaneously.
	
	\section{Summary and outlook}\label{sec:summary}
	Using IXPE X-ray polarization measurements of AGNs in the $2$--$8\,\mathrm{keV}$ energy range, we derived a new constraint on the isotropic $d=5$ SME coefficient in the photon sector.
	IXPE is the first dedicated X-ray polarimetry mission in more than 40 years and the first such mission that has measured X-ray polarization from an extragalactic source.
	While claims of gamma-ray polarization of GRBs have previously been made, those measurements suffer from large systematic uncertainties because the instruments were not calibrated for polarization measurements.
	These detections are also in some tension with the non-detection of gamma-ray polarization from GRBs with the dedicated GRB polarimeter POLAR.\cite{polar}
	While constraints derived from these measurements are in principle stronger than those presented here,\cite{kostelecky_mewes_2009,kostelecky_mewes_2013,grb_polarimetry} they are affected by these systematic uncertainties.
	The constraints presented here do not suffer from the same issue because IXPE has been carefully calibrated for polarimetry prior to launch.
	
	In a future paper, we will use the same data in combination with optical polarization measurements to constrain all SME photon-sector coefficients of $d=5$ and possibly higher mass dimension using the techniques described in Ref.~\refcite{friedman_etal_2020}.
	
	While IXPE polarization detections from additional extragalactic sources will continue to tighten possible constraints, another significant improvement in sensitivity is to be expected from polarimetry at higher energy enabled by proposed gamma-ray burst polarimetry missions or the upcoming Compton Spectrometer and Image (COSI) mission.
	
	\section*{Acknowledgments}
	This work is funded in part by NASA grants 80NSSC24K0205, 80NSSC24K0636, and 80NSSC24K1762.


\begin{thebibliography}{xx}
		
		\bibitem{datatables}
		\textit{Data Tables for Lorentz and CPT Violation}, V.A.\ Kosteleck\'y and N.\ Russell, 2025 edition, arXiv:0801.0287v18.
		
		\bibitem{liv}
		R.C.\ Myers and M.\ Pospelov, Phys.\ Rev.\ Lett.\ \textbf{90}, 211601 (2003);
		T.G.\ Rizzo, J.\ High Energy Phys.\ \textbf{2005}, 036 (2005);
		V.A.\ Kosteleck\'y and S.\ Samuel, Phys.\ Rev.\ D \textbf{39}, 683 (1989);
		C.P.\ Burgess \etal, J.\ High Energy Phys.\ \textbf{2002}, 043 (2002);
		R.\ Gambini and J.\ Pullin, Phys.\ Rev.\ D \textbf{59}, 124021 (1999);
		M.\ Pospelov and Y.\ Shang, Phys.\ Rev.\ D \textbf{85}, 105001 (2012);
		M.\ Li, Y.F.\ Cai, X.\ Wang and X.\ Zhang, Phys.\ Lett.\ B \textbf{680}, 118 (2009).
		
		\bibitem{dispersion}
		V.\ Vasileiou \etal, Phys.\ Rev.\ D \textbf{87}, 122001 (2013);
		F.\ Kislat and H.\ Krawczynski, Phys.\ Rev.\ D \textbf{92}, 045016 (2015);
		J.J.\ Wei and X.F.\ Wu, Front.\ Phys.\ \textbf{16}, 44300 (2021).
		
		\bibitem{kaaret_2004}
		P.\ Kaaret, Nature \textbf{427}, 287 (2004);
		
		\bibitem{kostelecky_mewes_2013}
		V.A.\ Kosteleck\'y and M.\ Mewes, Phys.\ Rev.\ Lett.\ \textbf{110}, 201601 (2013).
		
		\bibitem{kostelecky_mewes_2009}
		V.A.\ Kosteleck\'y and M.\ Mewes, Phys.\ Rev.\ D \textbf{80}, 015020 (2009).
		
		\bibitem{optical_polarimetry}
		F.\ Kislat and H.\ Krawczynski, Phys.\ Rev.\ D \textbf{95}, 083013 (2017);
		F.\ Kislat, Symmetry \textbf{10}, 596 (2018);
		R.\ Gerasimov, P.\ Bhoj, and F.\ Kislat, Symmetry \textbf{13}, 880 (2021).
		
		\bibitem{sme}
		D.\ Colladay and V.A.\ Kosteleck\'y, Phys.\ Rev.\ D \textbf{55}, 6760 (1997);
		D.\ Colladay and V.A.\ Kosteleck\'y, Phys.\ Rev.\ D \textbf{58}, 116002 (1998);
		V.A.\ Kosteleck\'y and M.\ Mewes, Phys.\ Rev.\ D \textbf{66}, 056005 (2002);
		V.A.\ Kosteleck\'y, Phys.\ Rev.\ D \textbf{69}, 105009 (2004).
		
		\bibitem{friedman_etal_2020}
		A.S.\ Friedman \etal, Phys.\ Rev.\ D \textbf{102}, 043008 (2020).
		
		\bibitem{weisskopf_etal_2022}
		M.\ Weisskopf \etal, J.\ Astron.\ Telescopes Instrum.\ Syst.\ \textbf{8}, 026002 (2022).
		
		\bibitem{digesu_etal_2022}
		L.\ Di Gesu \etal, Astrophys.\ J.\ Lett.\ \textbf{938}, L7 (2022).
		
		\bibitem{liodakis_etal_2022}
		I.\ Liodakis \etal, Nature \textbf{611}, 677 (2022).
		
		\bibitem{ursini_etal_2022}
		F.\ Ursini \etal, Mon.\ Not.\ Royal Astron.\ Soc.\ \textbf{519}, 50 (2023).
		
		\bibitem{marin_etal_2024}
		F.\ Marin \etal, Astron.\ Astrophys.\ \textbf{689}, A238 (2024).
		
		\bibitem{ngc4151}
		V.E.\ Gianolli \etal, Mon.\ Not.\ Royal Astron.\ Soc.\ \textbf{523}, 4468 (2023);
		V.E.\ Gianolli \etal, Astron.\ Astrophys.\ \textbf{691}, A29 (2024).
		
		\bibitem{banerjee_etal_2025}
		A.\ Banerjee \etal, arXiv:2504.12410.
		
		\bibitem{kouch_etal_2024}
		P.M.\ Kouch \etal, Astron.\ Astrophys.\ \textbf{689}, A119 (2024).
		
		\bibitem{ingram_etal_2023}
		A.\ Ingram \etal, Mon.\ Not.\ Royal Astron.\ Soc.\ \textbf{525}, 5437 (2023).
		
		\bibitem{errando_etal_2024}
		M.\ Errando \etal, Astrophys.\ J.\ \textbf{963}, 5 (2024).
		
		\bibitem{ehlert_etal_2023}
		S.R.\ Ehlert \etal, Astrophys.\ J.\ \textbf{959}, 61 (2023).
		
		\bibitem{middei_etal_2023}
		R.\ Middei \etal, Astrophys.\ J.\ Lett.\ \textbf{942}, L10 (2023).
		
		\bibitem{polar}
		M.\ Kole \etal, Astron.\ Astrophys.\ \textbf{644}, A124 (2020).
		
		\bibitem{grb_polarimetry}
		K.\ Toma \etal, Phys.\ Rev.\ Lett.\ \textbf{109}, 241104 (2012);
		P.\ Laurent \etal, Phys.\ Rev.\ D \textbf{83}, 121301(R) (2011);
		F.W.\ Stecker, Astropart.\ Phys.\ \textbf{35}, 95 (2011).
		
	\end{thebibliography}
\end{document}